\documentclass[10pt]{elsart}
\usepackage{epsfig}

\begin{document}

\begin{frontmatter}

{\rightline{CERN-TH/2003-198}}
{\rightline{hep-ph/xxxx}}

\title{\bf{Universality of Yukawa Couplings Confronts Recent Neutrino
Data}}

\author{G.C. Branco,\thanksref{gustavo}}
\author{M.N. Rebelo\thanksref{gui},}  
\author{and J.I. Silva-Marcos\thanksref{juca}}

\address{Departamento de F{\'\i}sica, Grupo Te{\'o}rico de F{\'\i}sica
de Part{\'\i}culas,\\
Instituto Superior T\'{e}cnico, Av. Rovisco Pais, 1049-001
Lisboa, Portugal.}

\thanks[gustavo]{On leave of absence at Theory Division, 
CERN, Geneva, Switzerland. E-mail address: gbranco@alfa.ist.utl.pt}
\thanks[gui]{On leave of absence at Theory 
Division, CERN, Geneva, Switzerland. E-mail address: 
rebelo@alfa.ist.utl.pt}
\thanks[juca]{E-mail address: Joaquim.Silva-Marcos@cern.ch}

\begin{abstract}
We propose a flavour structure for the leptonic sector
of the Standard Model, based on the idea of universality
of Yukawa couplings, which accommodates all the experimental
data on neutrino masses and mixing, at the same time
predicting specific correlations between low energy measurable 
quantities, such as the ratio of neutrino squared mass differences, 
$|U_{13}|$, the leptonic Dirac phase, and the double-beta
decay mass parameter. We also point out that 
it is possible, in this framework,
to generate a sufficient amount
of baryon asymmetry of the Universe through leptogenesis.
\end{abstract}
\end{frontmatter}

\section{Introduction}

The pattern of fermion masses and mixing remains a fascinating, unsolved
puzzle. Within the Standard Model (SM) fermion masses and mixing are
generated after gauge symmetry breaking; their flavour structure is
arbitrary, since no symmetry of the SM constrains the flavour dependence of
Yukawa couplings. The recent evidence for neutrino oscillations implies
non-vanishing neutrino masses, which in turn requires an extension of the SM.
Within the SM, neutrinos are strictly massless, since no Dirac masses are
possible because of the
absence of right-handed neutrinos, and no left-handed
Majorana masses can be generated in higher orders owing to exact B-L
conservation.

The simplest extension of the SM, which leads to non-vanishing neutrino
masses, consists of adding to the spectrum of the SM three right-handed
neutrinos (one per generation). If no other symmetries are introduced, one
is then naturally led to non-vanishing, but naturally small, neutrino masses
through the see-saw mechanism.

It was suggested some time ago that the diversity of quark masses and mixing
may be closely related to the fact that Yukawa couplings are the only
couplings of the SM that can be complex. In particular, it was shown that
the observed pattern of quark masses and mixing can be accounted for within
the framework of the
Universality of Strength of Yukawa couplings (USY), where
Yukawa couplings all have the same modulus, the flavour dependence being all
contained in the phases \cite{Branco:fj}. This leads to mass matrices of the
form: $M_q = c_q [\exp (i {\theta} _{jk}^q)] $. These mass matrices have
very interesting properties \cite{Kalinowski:1990mz}-- 
\cite{Branco:1996fb}, and it was shown \cite{Hung:2001hw}
that such a pattern can be generated in the context of extra dimensions with
fermions localized in a thick brane. The USY hypothesis was extended to the
leptonic sector in previous works, 
and in particular it was shown in \cite
{Branco:1999yf} that it is possible within USY to obtain the 
large mixing angle solar
solution (LMA)
together with large atmospheric mixing.

Constraints on the structure of the neutrino mass matrix and CP violation
and their implications can be analysed \cite{sf}, based solely on the current
data available from experiments with solar \cite{Fukuda:2002pe}, 
\cite{Ahmad:2002ka} and atmospheric \cite{Nakaya:2002ki}  neutrinos, 
reactor
experiments \cite{Apollonio:1999ae}, \cite{Eguchi:2002dm}, neutrinoless
double beta decay searches \cite{kkg}, astrophysics and cosmology \cite
{Bennett:2003bz}. If CP were conserved in the leptonic sector, all six of
the independent entries of the effective low energy 
neutrino mass $m_{\mathrm{eff}}$, 
resulting from the see-saw mechanism, could be determined from
experiment, through the measurement of three neutrino masses and three
mixing angles. If one assumes the general situation, one is led to leptonic
CP violation at low energies, with a $3\times 3$ complex symmetric effective
neutrino mass matrix, corresponding to six independent moduli and three
phases. It has been emphasized \cite{Frampton:2002yf} that this leads to a
situation where \textit{no currently conceivable set of feasable experiments
can fully determine the neutrino mass matrix}. This is because 
realistic experiments can only measure seven 
independent quantities, to wit $\Delta m_{12}^{2}$, 
$\Delta m_{23}^{2}$, $\theta _{12}$, $\theta _{13}$, 
$\theta _{23}$, $\delta $, $\mid \langle m \rangle \mid $ 
(i.e. two squared neutrino mass
differences, three mixing angles, 
the Dirac-type CP violating phase and the
effective mass element relevant to neutrinoless 
double beta decay, which is
sensitive to the presence of Majorana phases). This observation led to
various suggestions to limit the number of parameters in the neutrino mass
matrix, such as the possibility that the neutrino mass matrix has texture
zeros \cite{Frampton:2002yf}, \cite{varios}. An alternative proposal, to
remove the ambiguities in the reconstruction of the neutrino mass matrix,
consists of imposing the weak basis 
(WB)-independent condition $\mbox{det}\,(m_{\mathrm{eff}})=0$, 
without imposing texture zeros \cite{zerodet}. In
this paper we apply the USY hypothesis together with the condition that the
determinant of $m_{\mathrm{eff}}$ is zero, to construct a highly predictive
scheme where not only the low energy neutrino mass matrix can be
reconstructed, but we also obtain specific predictions both for low energy
phenomena and for leptogenesis \cite{Fukugita:1986hr}.

\section{Framework}

\subsection{Choice of USY ansatz}

We add to the spectrum of the SM three right-handed neutrino fields $\nu
_{R}^{0}$ (one per generation), which leads, after spontaneous symmetry
breaking, to the following mass terms: 
\begin{eqnarray}
\mathcal{L}_{m} &=&-\left[ \overline{{\nu }_{L}^{0}}\ m\ \nu _{R}^{0}+\frac{1%
}{2}\nu _{R}^{0T}\ C\ M\ \nu _{R}^{0}+\overline{l_{L}^{0}}\ m_{l}\
l_{R}^{0}\right] +\mathrm{h.c.}=  \nonumber \\
&=&-\left[ \frac{1}{2}n_{L}^{T}\ C\ \mathcal{M}^{*}n_{L}+\overline{l_{L}^{0}}%
\ m_{l}\ l_{R}^{0}\right] +\mathrm{h.c.} \; ,  \label{lm}
\end{eqnarray}
where $m$, $M$ and $m_{l}$ denote the neutrino Dirac mass matrix, the
right-handed neutrino Majorana mass matrix and the charged lepton mass
matrix, respectively, $n_{L}=({\nu }_{L}^{0},{(\nu _{R}^{0})}^{c})$ (written
here as a line instead of a column in order to save space) and $\mathcal{M}$
is given by 
\begin{equation}
\mathcal{M}=\left( 
\begin{array}{cc}
0 & m \\ 
m^{T} & M
\end{array}
\right) \; .
\end{equation}
The zero entry in $\mathcal{M}$ reflects the absence of a term of the form 
$\frac{1}{2}{\nu }_{L}^{0T}\ C\ m_{L}\ \nu _{L}^{0}$, 
which requires, in order to be
generated at tree level, an 
extension of the Higgs sector of the SM.
The matrix $M$ corresponds to an $SU(2)\times U(1)\times SU(3)_{c}$
invariant mass term; therefore its entries can naturally be of a scale
much higher than the electroweak scale. Under this assumption, the light
neutrino masses are obtained, to an excellent approximation, from the
diagonalization of the effective neutrino mass matrix defined as $m_{\mathrm{%
eff}}\equiv -m\ M^{-1}m^{T}$: 
\begin{equation}
-V^{\dagger }m\ M^{-1}m^{T}V^{*}=d_{\nu }\; .  \label{vmv}
\end{equation}
In this work we suggest the following specific USY ansatz for the three
different mass matrices appearing in $\mathcal{L}_{m}$: 
\begin{equation}
\begin{array}{ll}
m_{l}=c_{l}\ K\cdot \left[ 
\begin{array}{lll}
1 & 1 & 1 \\ 
1 & e^{ia_{l}} & 1 \\ 
1 & 1 & e^{ib_{l}}
\end{array}
\right], &  \quad m=c_{D}\ \left[ 
\begin{array}{lll}
1 & 1 & 1 \\ 
1 & 1 & e^{ia} \\ 
1 & 1 & e^{ib}
\end{array}
\right], \\ 
&  \\ 
M=c_{R}\left[ 
\begin{array}{lll}
1 & 1 & 1 \\ 
1 & e^{iA} & 1 \\ 
1 & 1 & e^{iB}
\end{array}
\right], & 
\end{array}
\label{mmm}
\end{equation}
where $K\equiv \mathrm{diag}(1,1,e^{i\theta })$ 
and $c_{l}$, $c_{D}$, and $c_{R}$ are real constants, 
i.e. we assume that it is possible to cast all
of the leptonic mass matrices simultaneously in a USY form. In order to
implement the condition $\mbox{det}\,(m_{\mathrm{eff}})=0$, we have chosen
the Dirac mass matrix $m$ to have the USY form, but with vanishing
determinant. Within the USY framework, all inequivalent forms leading to a
vanishing determinant have been classified, and it was shown that there are
only two classes of solutions \cite{Branco:1995pw}. The Majorana mass matrix
is also of the USY form and it has the nice feature that its inverse has a
simple form. From Eq. (\ref{mmm}) one obtains: 
\begin{equation}
M^{-1}=\frac{1}{N}\left[ 
\begin{array}{ccc}
e^{i(A+B)}-1 & 1-e^{iB} & 1-e^{iA} \\ 
1-e^{iB} & e^{iB}-1 & 0 \\ 
1-e^{iA} & 0 & e^{iA}-1
\end{array}
\right] ;\quad N=c_{R}(e^{iA}-1)(e^{iB}-1) \; . \label{imn}
\end{equation}
It is useful to write $m$ as: 
\begin{equation}
m=c_{D}(\Delta +P)\; ,  \label{cdp}
\end{equation}
with 
\begin{equation}
P=\left[ 
\begin{array}{ccc}
0 & 0 & 0 \\ 
0 & 0 & e^{ia}-1 \\ 
0 & 0 & e^{ib}-1
\end{array}
\right] \qquad \mbox{and}\qquad \Delta =\left[ 
\begin{array}{ccc}
1 & 1 & 1 \\ 
1 & 1 & 1 \\ 
1 & 1 & 1
\end{array}
\right]\; .  \label{pde}
\end{equation}
From Eqs. (\ref{imn}), (\ref{cdp}) and (\ref{pde}), 
and using the fact that $\Delta M^{-1}P+PM^{-1}\Delta =0$, 
we obtain: 
\begin{equation}
-m_{\mathrm{eff}}\equiv m\frac{1}{M}m^{T}=c_{\mathrm{eff}}\left[ 
\begin{array}{ccc}
1 & 1 & 1 \\ 
1 & 1+kx^{2} & kxy \\ 
1 & 1+kxy & 1+ky^{2}
\end{array}
\right]\; ,  \label{eff}
\end{equation}
with $k=\frac{1}{e^{iB}-1}$, $x=e^{ia}-1$, 
$y=e^{ib}-1$ and $c_{\mathrm{eff}}=c_{D}^{2}/c_{R}$. 
Notice that the parameter $A$ does not appear at low
energies. In the following we show 
that $a,b\ll 1$, so that $m_{\mathrm{eff}}$
is a small perturbation around $\Delta $.

The leptonic charged current interactions can then be written as: 
\begin{equation}
\mathcal{L}_{W}=-\frac{g}{\sqrt{2}}\ \overline{l_{L}\ }\gamma _{\mu }U\ {\nu 
}_{L}\ W^{\mu }+\mathrm{h.c.}\; ,  \label{cci}
\end{equation}
where the Pontecorvo--Maki--Nakagawa--Sakata (PMNS) 
matrix \cite{matrix} is
given by $U=V_{l}^{\dagger }V$, with $V_{l}$ denoting the unitary matrix
entering in the diagonalization of the charged lepton mass matrix: 
\begin{equation}
V_{l}^{\dagger }m_{d}m_{d}^{\dagger }V_{l}=d_{l}^{2}\; .  \label{vmd}
\end{equation}
We use the standard parametrization \cite{Hagiwara:fs} 
for $U$, factoring out the Majorana type phase 
\begin{equation}
U=\left( 
\begin{array}{ccc}
c_{12}c_{13} & s_{12}c_{13} & s_{13}e^{-i\delta } \\ 
-s_{12}c_{23}-c_{12}s_{23}s_{13}e^{i\delta } & \quad
c_{12}c_{23}-s_{12}s_{23}s_{13}e^{i\delta }\quad & s_{23}c_{13} \\ 
s_{12}s_{23}-c_{12}c_{23}s_{13}e^{i\delta } & 
-c_{12}s_{23}-s_{12}c_{23}s_{13}e^{i\delta } & c_{23}c_{13}
\end{array}
\right) \,\cdot P,  \label{pdg}
\end{equation}
where $c_{ij}\equiv \cos \theta _{ij}\ ,\ s_{ij}\equiv \sin \theta _{ij}\ $,
with all $\theta _{ij}$ in the first quadrant and 
$P=\mathrm{diag\ }(1,1,e^{i\alpha })$; $\delta $ 
is a Dirac-type phase (analogous to the one
of the quark sector) and $\alpha $ is the physical phase associated with the
Majorana character of neutrinos. In general, for three generations, there
are two Majorana phases, but the fact that one of the light neutrino masses
vanishes eliminates one of these phases.

\subsection{Counting of parameters}

In order to evaluate the predictive power of our ansatz, let us compare its
number of parameters with the total number of parameters present in general
with one zero  neutrino mass. We will do this comparison first taking
into account the full parameter space, and later considering only the
parameters that are relevant in the low energy limit.

\subsubsection{The full parameter space}

From Eq. (\ref{mmm}) one sees that our ansatz is characterized by ten
parameters namely $(c_{R},A,B)$, $(c_{D},a,b)$, $(c_{l},\theta
,a_{l},b_{l}) $. Let us now count the number of independent parameters
characterizing the lepton masses and mixing in the general case, but with
the implicit assumption that one light neutrino mass vanishes. The total
number of parameters can be obtained by going to the WB where both $m_{l}$
and $M$ are real, diagonal. Then apart from the six real parameters
characterizing the eigenvalues of $m_{l}$ and $M$, one has to take into
account the number of parameters needed for specifying, in the above WB, the
Dirac neutrino mass matrix $m$, taking into account the vanishing of its
determinant. In this case, it can be shown that any matrix $m$ can be
written as: 
\begin{equation}
m=\hat{U_{L}}\ d\ P_{1}\hat{U_{R}}\ P_{2}  \; , \label{wht}
\end{equation}
where $\hat{U_{L}}(\delta _{L})$ $\hat{U_{R}}(\delta _{R})$ are unitary
matrices containing only one phase each (\` a la 
Kobayashi--Maskawa) and 
\begin{equation}
d=\mathrm{diag}(0,m_{2},m_{3});\ \ P_{1}=\mathrm{diag}(1,e^{i\phi },1);\ \
P_{2}=\mathrm{diag}(e^{i\sigma _{1}},e^{i\sigma _{2}},1).  \label{dpp}
\end{equation}
We have already omitted the three phases that can always be rephased away
irrespective of the value of $\mbox{det}\, (m)$. 
From Eqs. (\ref{wht}) and (\ref{dpp})
it follows that $m$ is characterized by thirteen parameters, consisting of
eight real numbers and five phases. Together with the six eigenvalues of $%
m_{l}$ and $M$, one has altogether nineteen parameters needed to specify the
three leptonic mass matrices $m$, $m_{l}$ and $M$. This equals the number of
physical quantities (eight non-zero masses, six mixing angles, five CP
violating phases). The condition $\mbox{det}\,(m_{\mathrm{eff}})=0$
eliminates one mass and one phase, so that in general with see-saw one would
have twenty one parameters \cite{Endoh:2000hc}. 
This is to be compared with the
ten parameters characterizing our ansatz, thus showing its high predictive
power. In general, the phases appearing in $m$ generate  CP violation 
both at low and at high energies, 
where they are essential to
produce viable leptogenesis. Furthermore, it has been pointed out that these
CP violating phases may manifest themselves at low energies even without CP
violation at high energies \cite{Branco:2001pq}; conversely it is
possible to have CP violation at high energies with no CP violation at low
energies \cite{Rebelo:2002wj} either of Dirac or of Majorana type. Although
the above counting is useful in gauging the predictive power of our
ansatz, it is clear that at present we can only expect to be able to measure
low energy parameters. Therefore, it is interesting to make an analogous
counting, taking only those into account, as well as to discuss how the low
energy parameters are fixed.

\subsubsection{The low energy limit and fixing the parameters}

At low energies our ansatz is characterized by eight parameters, namely, $%
\theta $, $c_{l}$, $a_{l}$, $b_{l}$, $c_{\mathrm{eff}}$, $a$, $b$, $B$,
whilst ten parameters are needed to parametrize low energy data, namely $%
m_{e}$, $m_{\mu }$, $m_{\tau }$, and $m_{2}$, $m_{3}$ 
(the two non-vanishing
light neutrino masses), and also $\theta _{12}$, $\theta _{13}$, $\theta
_{23}$, $\delta $, and $\alpha $, which characterize the PMNS matrix.

The three independent real parameters $c_{l}$, $a_{l}$ and $b_{l}$ in $m_{l}$
are fixed once we impose the correct values for the charged lepton masses.
Taking into account the charged lepton mass hierarchy, we find the following
approximate relations \cite{Akhmedov:2000yt}: 
\begin{equation}
|a_{l}|\simeq 6\frac{m_{e}}{m_{\tau }}\,,~\quad \quad |b_{l}|\simeq \frac{9}{%
2}\frac{m_{\mu }}{m_{\tau }}\,,
\quad \quad |c_{l}|\simeq \frac{m_{\tau }}{3} \; .
\label{abc}
\end{equation}
With all four parameters of $m_{l}$ fixed, the diagonalizing matrix $V_{l}$
is then determined \cite{Akhmedov:2000yt}: 
\begin{equation}
\begin{array}{l}
V_{l}=K\cdot F\cdot W \\ 
\\ 
K\equiv \mathrm{diag}(1,1,e^{i\theta })
\end{array}
\quad ;\quad F=\left( 
\begin{array}{ccc}
~~\frac{1}{\sqrt{2}} & ~~\frac{1}{\sqrt{6}} & ~~\frac{1}{\sqrt{3}} \\ 
-\frac{1}{\sqrt{2}} & ~~\frac{1}{\sqrt{6}} & ~~\frac{1}{\sqrt{3}} \\ 
~0 & -\frac{2}{\sqrt{6}} & ~~\frac{1}{\sqrt{3}}
\end{array}
\right)\; , \quad  \label{fw}
\end{equation}
where for $W$ one has, to leading order: 
\begin{equation}
W\simeq {\left( 
\begin{array}{ccc}
\vspace*{0.15cm}1 & \frac{m_{e}}{\sqrt{3} m_{\mu }} & -i\sqrt{\frac{2}{3}}%
\frac{m_{e}}{m_{\tau }} \\ 
\vspace*{0.15cm}-\frac{m_{e}}{\sqrt{3}m_{\mu }} & 1 & i\frac{m_{\mu }}{\sqrt{%
2}m_{\tau }} \\ 
-i\sqrt{\frac{3}{2}}\frac{m_{e}}{m_{\tau }} & i\frac{m_{\mu }}{\sqrt{2}
m_{\tau }} & 1
\end{array}
\right) }\; .  \label{www}
\end{equation}

The four parameters $a$, $b$, $B$ and $c_{\mathrm{eff}}\equiv
c_{D}^{2}/c_{R} $ appearing in the effective neutrino mass matrix are
directly related to the light neutrino squared mass differences. We obtain
for the invariants (taking into account that $m_{1}=0$) $\chi \equiv \chi
(h)=m_{2}^{2}m_{3}^{2} $ and the trace $t\equiv \mathrm{Tr}%
(h)=m_{2}^{2}+m_{3}^{2}$ of the hermitian matrix $h\equiv m_{\mathrm{eff}}%
\mathrm{\ }m_{\mathrm{eff}}^{\dagger }$, expressed as functions of $a$, $b$, 
$B$ and $c_{\mathrm{eff}}$ :
\begin{equation}
\begin{array}{l}
\frac{\chi }{c_{\mathrm{eff}}^{4}}\ \ =4\ [\sin ^{2}(\frac{a}{2})+\sin ^{2}(%
\frac{b}{2})+\sin ^{2}(\frac{a-b}{2})]^{2}\ /\ \sin ^{2}(\frac{B}{2}) \\ 
\\ 
\frac{t\ }{c_{\mathrm{eff}}^{2}}=9+[3+\cos (a-b)+\cos (a+b-B)+\cos (a-b)\cos
(a+b-B) \\ 
\ +2\cos (B)-2\cos (a)-2\cos (b)-2\cos (a-B)-2\cos (b-B)]/\ \sin ^{2}(\frac{B%
}{2}) \; .
\end{array}
\label{veel}
\end{equation}
Thus, assuming small values for $a$, $b$ and $B$, one finds the approximate
relations: 
\begin{eqnarray}
\chi &\simeq &c_{\mathrm{eff}}^{4}\ \frac{4}{B^{2}}(a^{2}+b^{2}-ab)^{2},
\label{bab} \\
t &\simeq &9\ c_{\mathrm{eff}}^{2}\; ,  \qquad \qquad 
|c_{\mathrm{eff}}|\simeq m_{3}/3
\label{tchi}
\end{eqnarray}
and for the neutrino mass ratio $r\equiv
\frac{m_{2}} {m_{3}}$:
\begin{equation}
\frac{\sqrt{\chi }}{t}=\frac{r}{1+r}\simeq r\simeq \sqrt{\frac{\Delta m_{%
\mathrm{sol}}^{2}}{\Delta m_{\mathrm{atm}}^{2}}}\simeq \frac{2}{9B}%
(a^{2}+b^{2}-ab)\; .  \label{quo}
\end{equation}
Note that for a fixed mass ratio $r$ and for a fixed $B$, this formula
constrains the values of $a$ and $b$ to an (almost) ellipse.

It is also possible to obtain simple approximate expressions for $\theta _{%
\mathrm{atm}}$ and $\theta _{\mathrm{sol}}$. Assuming $a$ and $b$ ``small'',
$\sin ^{2}(2\theta _{\mathrm{atm}})$ is, to leading order,
only a function of $\theta $, and one obtains: 
\begin{equation}
\sin ^{2}(2\theta _{\mathrm{atm}})\simeq 4\cdot p\ (1-p)\quad ;\quad p=\frac{%
8}{9}\sin ^{2}\left( \frac{\theta }{2} \right) \; ,  \label{atm}
\end{equation}
whilst $\sin ^{2}(2\theta _{\mathrm{sol}})$ is approximately given by: 
\begin{equation}
\sin ^{2}(2\theta _{\mathrm{sol}})\simeq 4\cdot q\ (1-q)\quad ;\quad q=\frac{%
3a^{2}}{4(a^{2}+b^{2}-ab)}\; .  \label{sin}
\end{equation}
It is clear from Eqs. (\ref{quo}) and (\ref{sin}) that, in the range of
validity of these expressions, physics at low energies only puts constraints
on the ratios $a/\sqrt{B}$, $b/\sqrt{B}$, and not on $a$, $b$ and $B$
separately. Another interesting point is the fact that for $\theta _{\mathrm{%
sol}}=30^{\circ }$, Eq. (\ref{sin}) has two very simple solutions of the
form $b=2a$ and $b=-a$, corresponding to straight lines in the $(a,b)$
plane. In the next section the currently allowed ranges for $\Delta
m_{21}^{2}$, $\Delta m_{32}^{2}$, $\tan ^{2}(\theta _{12})$ ($\theta
_{12}\equiv \theta _{\mathrm{sol}}$) and $\tan ^{2}(\theta _{23})$ ($\theta
_{23}\equiv \theta _{\mathrm{atm}}$) are given; it can be seen that this
value for $\theta _{\mathrm{sol}}$ is close to the central allowed
experimental value. The numerical search for the allowed parameter space was
performed with exact expressions and led to regions in the ($a$, $b$) 
plane in the neighbourhood of the above straight lines.

For hierarchical heavy neutrino masses, it is clear that $A$ and $B$ are
small and verify relations similar to those for $a_{l}$ and $b_{l}$: 
\begin{equation}
|A|\simeq 6\frac{M_{1}}{M_{3}}\,,~\quad \quad |B|\simeq \frac{9}{2}\frac{%
M_{2}}{M_{3}}\,,\quad \quad A,B\ll 1\; .  \label{abg}
\end{equation}
The fact that one of the light neutrino masses is $0$ constrains 
$c_{\mathrm{eff}}$ to be of the order of
$\sqrt{\Delta m_{\mathrm{atm}}^{2}}\sim
10^{-2}$ eV. Taking $c_{D}$ to be of the electroweak scale ($\sim 10^{11}
\ \mbox{eV}$) one obtains $c_{R}$ of order $\sim 10^{16}$ GeV, 
implying $M_{3}$ of the same order. Heavy neutrino masses are 
not further constrained by experiment. The
value of the parameter $A$ does not affect 
low energy physics, since it does
not appear in the exact expression for 
$m_{\mathrm{eff}}$, given in Eq. (\ref{eff}). 
In this framework, neither $|U_{13}|$ nor $\delta $ and $\alpha $
are free parameters. It is possible to have $|U_{13}|$ close to its
experimental limit and the CP violating parameter $I$ defined by 
$|\mathrm{Im}\left[ U_{ij}U_{kl}U_{kj}^{*}U_{il}^{*}\right] |$, 
which is sensitive to the
value of $\delta $, can be of order $10^{-2}$, within the reach of future
neutrino experiments, meaning that in our framework the phase $\delta $ can
be large.

In the next section we present the implications of our model for low energy
physics.

\section{Examples and predictions}

It was shown in section 2 that our USY model has fewer free parameters than
the total number of measurable quantities at low energies. Therefore, the
model predicts specific correlations among these quantities. In this
section, we first describe how the analysis was performed and present a
specific interesting example, which is taken up again in the discussion of
leptogenesis. Next, we present, in the form of various plots, correlations
between physical quantities, which are implied by our framework,
and comment on the main features of our model.

\subsection{Strategy}

In our analysis, illustrated in part by the graphs that follow, we have
fixed the parameters $c_{l}$, $a_{l}$ and $b_{l}$ in such a way that the PDG
values \cite{Hagiwara:fs} given for the charged lepton masses are obtained.
The masses of $m_{e}=0.511$ MeV, $m_{\mu }=106$ MeV 
and $m_{\tau }=1777$ MeV correspond to $a_{l}=1.725089\times 10^{-3}$, 
$b_{l}=0.26785$ and $c_{l}=593.3863$ MeV. 
With these values the leptonic mixing matrix is
obtained, still with $\theta $ as a free parameter 
(since factorizable phases
do not affect the eigenvalues of the mass matrix). 
The resulting matrix $V_{l}$ is given by: {\footnotesize 
\begin{equation}
V_{l}=K\cdot \left[ 
\begin{array}{ccc}
0.70824 & -0.40554 & 0.57787 \\ 
-0.70597-0.00031i & -0.40949-0.00006i & 0.57786+0.00033i \\ 
-0.00227-0.00030i & 0.81400+0.07250i & 0.57404+0.05125i
\end{array}
\right] .  \label{cmm}
\end{equation}
} As expected, this matrix is very close to the matrix $F$ in Eq. (\ref{fw}%
). \footnote{Five decimal digits are quite sufficient 
for the computation of the matrix $U $, which is 
not yet known experimentally with such high precision. The
check of the eigenvalue equation for the charged lepton masses 
requires that we know $V_{l}$ with much higher precision 
since the squared masses of the charged leptons differ by
many orders of magnitude.} At low
energies, the remaining relevant parameters are 
$c_{\mathrm{eff}}$, $a,b$, $B$ and $\theta $. 
In our search for allowed regions of parameters we have
scanned $\theta $ from $0$ to $2\pi $, $a$ and $b$ in the region $[-0.4, 
0.4]$, and $B$ in the interval $[-0.021, 0.021]$. 
Notice that $B$ is chosen to
be much smaller than $1$, so that heavy neutrinos have hierarchical masses
(these also depend on $A$). It follows from Eq. (\ref{quo}) that, in this
case, the values of $a$ and $b$ compatible with the experimental constraints
on neutrino masses are well inside the interval given above for these
parameters. For each fixed set of parameters, the ratio $\Delta m_{21}^{2}$/$%
\Delta m_{32}^{2}$ as well as $\tan ^{2}(\theta _{12})$ and $\tan
^{2}(\theta _{23})$ were determined and compared 
with the currently allowed
experimental ranges, consistent with the first results from K2K and KamLAND,
as given in \cite{Gonzalez-Garcia:2003qf} (notice that $\theta _{12}$ is the
solar angle and $\theta _{23}$ the atmospheric angle): 
\begin{eqnarray}
(1.5)\ 2.2 &<&\Delta m_{32}^{2}/10^{-3}\ \mbox{eV}^{2}<3.0\ (3.9)  
\nonumber \\
(0.45)\ 0.75 &<&\tan ^{2}(\theta _{23})<1.3\ (2.3)  \nonumber \\
(5.4)\ 6.7 &<&\Delta m_{21}^{2}/10^{-5}\ \mbox{eV}^{2}<7.7\ (10.0)\qquad %
\mbox{and}  \label{k2k} \\
(14.0) &<&\Delta m_{21}^{2}/10^{-5}\ \mbox{eV}^{2}<(19.0)  \nonumber \\
(0.29)\ 0.39 &<&\tan ^{2}(\theta _{12})<0.51\ (0.82) \; .  \nonumber
\end{eqnarray}
These are  $1\sigma $ ( $3\sigma $) CL intervals; the second range of $%
\Delta m_{21}^{2}$ corresponds
to solutions in the upper LMA island (at
present the results of the solar and KamLAND analyses still allow for an
ambiguity in the determination of $\Delta m_{21}^{2}$ 
at CL $>$  $2.5\sigma $). Furthermore 
in our programme we have imposed a cut in $|U_{13}|^{2}\equiv
\sin ^{2}(\theta _{13})<0.02$. We have used  $1\sigma $ CL intervals for the
solar and atmospheric angles; for the mass ratio we worked with a  $3\sigma $
CL, since it is a feature of our model that solutions with $|U_{13}|$ close
to the experimental bound and leading to a large value of $|I|$ correspond
to values of $\Delta m_{21}^{2}$ below 
$6.7\times 10^{-5} \ \mbox{eV}^{2}$. Our
ansatz predicts that sizable CP violation in neutrino oscillations is not
compatible with the upper LMA.

The numerical analysis was performed with exact formulae. Since we constrain
the ratio of masses rather than the masses 
themselves, we do not need to fix $%
c_{\mathrm{eff}}$. Let us, for the sake of illustration, present a possible
solution in the region $b\simeq 2a$, with both $a$ and $b$ positive: 
\begin{equation}
a=0.058998;\ b=0.102998;\ \theta =1.9\ \mbox{rad} \; .
\end{equation}
In this example we obtain, fixing $c_{\mathrm{eff}}=0.0163$ eV: 
\begin{eqnarray}
\Delta m_{21}^{2} &=&6.6\times 10^{-5}\ \mbox{eV}^{2},\ \Delta
m_{32}^{2}=2.7\times 10^{-3}\ \mbox{eV}^{2},  \nonumber \\
\ \tan ^{2}(\theta _{12}) &=&0.46,\ \tan ^{2}(\theta _{23})=0.99 \\
|U_{13}| &=&0.13\ \mbox{and}\ |I|=0.02 \; . \nonumber  \label{xpl}
\end{eqnarray}
This example has large $|U_{13}|$, still compatible with the experimental
bound \cite{Gonzalez-Garcia:2003qf} of $\sin ^{2}(\theta _{13})<0.02\
(0.052) $ at  $1\sigma $ ( $3\sigma $) and large $|I|$, within the reach of
the neutrino experiments being planned. This is a generic
characteristic of the examples in the region around $b\simeq 2a$ and $\theta 
$ close to 1.9 rad. This example is taken again in the discussion of
leptogenesis.

\subsection{Correlations and implications for low energy CP violation}

The simplest way of describing the predictions of our USY model is by
presenting the correlations implied by it, for various physical quantities.
Our results are depicted in Figs. 1--5. 
The points correspond to solutions
satisfying all the experimental constraints. We have chosen to depict
correlations between physical observables rather than between the initial
parameters of the Lagrangian. There are two different disconnected regions
for $|U_{13}|^2$, once all other experimental constraints are imposed, one
of them characterized by very small values of $|U_{13}|^2$, which are more
than one order of magnitude below its upper bound, obtained around $b \simeq
-a$, and the other around $b \simeq 2a$, with values of the order of magnitude
of the upper bound. Only in the region of larger $|U_{13}|^2$ can one expect
to observe CP violation at the neutrino factories being planned at present.

In Fig. 1 we present the correlation between $|U_{13}|^{2}$ and $\tan
^{2}(2\theta _{\mathrm{sol}})$. It is clear that there are two regions
corresponding respectively to small and relatively large values of $%
|U_{13}|^{2}$, as mentioned above. However, in each one of these regions, $%
\tan ^{2}(2\theta _{\mathrm{sol}})$ may have any value within the
experimental bound. In Fig. 2, we present the correlation between $r\equiv
(\Delta m_{21}^{2}$/$\Delta m_{32}^{2})^{1/2}$ and $|U_{13}|^{2}$. Again, we
find two regions corresponding to smaller and larger values of $|U_{13}|^{2}$%
. Furthermore, Fig. 2 clearly shows that in the present model, relatively
large values of $|U_{13}|^{2}$ are only possible for small values of the
mass ratio $r$. In Fig.~3 we plot the correlation between $|U_{13}|^{2}$ and
the CP-odd rephasing invariant $I$. In terms of mixing angles, $I$ is given
by: 
\begin{equation}
I=\frac{1}{8}\sin (2\theta _{12})\sin (2\theta _{23})\sin (2\theta
_{13})\cos (\theta _{13})\sin \delta  \; ; \label{jsc}
\end{equation}
from this expression it follows that a ``large'' value of $I$ (e.g. $I\sim
10^{-2}$) requires $|U_{13}|$ close to its upper bound and an unsuppressed
Dirac phase $\delta $. It can be clearly seen that there are many solutions
corresponding to a large Dirac phase. For $I$ of the order of $10^{-2}$,
leptonic CP violation may be detected at neutrino factories through the
study of neutrino oscillations. In Fig. 4 we present the allowed regions of $%
I$ versus $r$. It is clear that relatively large values of $I$ are only
possible for small values of $r$. This result was to be expected, since
large values $|U_{13}|$ are only obtained for regions of small $r$.

\subsection{Implications for double beta decay}

In Fig. 5 the correlation between $(\delta -\alpha )$ (the phase appearing
in $U_{13}$ as defined in Eq. (\ref{pdg})), and $|U_{13}|^{2}$ is given.
Neutrinoless double beta decay measures $|\langle m \rangle | $ defined by: 
\begin{equation}
|\langle m \rangle |\equiv 
|m_{1}U_{11}^{2}+m_{2}U_{12}^{2}+m_{3}U_{13}^{2}|=|m_{2}\
s_{12}^{2}c_{13}^{2}+m_{3}\ s_{13}^{2}\ e^{2i(\delta -\alpha )}| \; , 
\label{mid}
\end{equation}
where we have used in the second equality the fact that in our ansatz $m_{1}$
vanishes. It is clear that this observable is sensitive to the combination
of phases $(\delta -\alpha )$. In models with $m_{1}=0$, the term $%
m_{2}s_{12}^{2}c_{13}^{2}$ is of order $10^{-3}$ eV; on the other hand the
product $m_{3}s_{13}^{2}$ can only reach this order of magnitude for $%
s_{13}^{2}$ around its maximal experimental bound of order $10^{-2}$. In
this case, cancellations might occur for $(\delta -\alpha )$ close to $\pi
/2 $ rad. Yet Fig. 5 shows this phase to be somewhat below $\pi /4$ rad in
the region of larger $|U_{13}|^{2}$, thus rendering cancellations
impossible. As a result our framework predicts 
$|\langle m \rangle |  $ of order $10^{-3} \ \mbox{eV}$,  
which is somewhat below the range favoured by the
Moscow--Heidelberg experiment \cite{kkg}, 
but within the reach of the next
generation of experiments.

\section{Leptogenesis in this framework}

It is well known that, in the case of hierarchical heavy neutrinos, the
baryon asymmetry generated through thermal leptogenesis only depends on four
parameters \cite{Buchmuller}: the mass $M_{1}$ of the lightest heavy
neutrino, together with the corresponding CP asymmetry $\epsilon _{N_{1}}$ in
their decays, as well as the effective neutrino mass $\widetilde{m_{1}}$
defined as 
\begin{equation}
\widetilde{m_{1}}=(m^{\dagger }m)_{11}/M_{1}  \label{mtil}
\end{equation}
in the weak basis where $M$ is diagonal, real and positive and, finally, the
sum of all light neutrino masses squared, ${\overline{m}}%
^{2}=m_{1}^{2}+m_{2}^{2}+m_{3}^{2}$, which controls an important class of
washout processes.

The CP asymmetry $\epsilon _{N_{1}} $ generated by the lightest heavy
neutrino is explicitly given by \cite{Covietal}: 
\begin{equation}
\epsilon _{N_{1}}=\frac{1}{8\,\pi \,v^{2}} \frac{1}{({m}^{\dag }m)_{11}}%
\,\sum_{i=2,3}\mathrm{Im}\left[ ({m}^{\dag }m)_{1i}^{2}\right] \left[
f\!\left( \frac{M_{i}^{2}}{M_{1}^{2}}\right) +g\!\left( \frac{M_{i}^{2}}{%
M_{1}^{2}}\right) \right] \ ,  \label{lepto1}
\end{equation}
where $v=\langle \phi ^{0}\rangle /\sqrt{2}\simeq 174\,$GeV and $f(x)$, $%
g(x) $ denote the one-loop vertex and self-energy corrections: 
\begin{equation}
f(x)=\sqrt{x}\left[ 1+(1+x)\ln \left( \frac{x}{1+x}\right) \right] \quad
,\quad g(x)=\frac{\sqrt{x}}{1-x}\,.  \label{lepto2}
\end{equation}
In the limit $M_{1}\ll M_{2},M_{3}$, the CP asymmetry (\ref{lepto1}) can be
written as: 
\begin{equation}
\epsilon _{N_{1}}\simeq -\frac{3}{16\,\pi v^{2}}\,\left( I_{12}\,\frac{M_{1}%
}{M_{2}}+I_{13}\,\frac{M_{1}}{M_{3}}\right) \,,  \label{lepto3}
\end{equation}
where 
\begin{equation}
I_{1i}\equiv \frac{\mathrm{Im}\left[ (m^{\dagger }m)_{1i}^{2}\right] }{%
(m^{\dagger }\,m)_{11}}\ .  \label{lepto4}
\end{equation}
The lepton asymmetry $Y_{L}$ is connected to the CP asymmetry through the
relation \cite{Kolb:vq} 
\begin{equation}
Y_{L}=\frac{n_{L}-n_{\bar{L}}}{s}=d\,\frac{\epsilon _{N_{1}}}{g_{*}}\; ,
\label{lepto5}
\end{equation}
where $g_{*}$ is the effective number of relativistic degrees of freedom
contributing to the entropy and $d$ is the 
so-called dilution factor, which
accounts for the washout processes (inverse decay and lepton number
violating scattering). In the SM case, $g_{*}=106.75$.

The produced lepton asymmetry $Y_{L}$ is converted into a net baryon
asymmetry $Y_{B}$ through the $(B+L)$-violating sphaleron processes \cite
{Kuzmin:1985mm}. One finds the relation \cite{Khlebnikov:sr}, \cite
{Harvey:1990qw}:
\begin{equation}
Y_{B}=\xi \,Y_{B-L}=\frac{\xi }{\xi -1}\,Y_{L}\;\;,\;\;\xi = \frac{%
8\,N_{f}+4\,N_{H}}{22\,N_{f}+13\,N_{H}}\,,  \label{lepto6}
\end{equation}
where $N_{f}$ and $N_{H}$ are the number of fermion families and complex
Higgs doublets, respectively. Taking into account that $N_{f}=3$ and $%
N_{H}=1 $ for the SM, we get $\xi \simeq 1/3$ and 
\begin{equation}
Y_{B}\simeq -\frac{1}{2}\,Y_{L}\,.  \label{lepto7}
\end{equation}

Successful leptogenesis would require $\epsilon _{N_{1}} $ of order $10^{-8}$
if washout processes could be neglected (i.e. $d=1$), in order to reproduce
the observed ratio of baryons to photons, which is given by \cite
{Bennett:2003bz}: 
\begin{equation}
\frac{n_{B}}{n_{\gamma}}= (6.1 ^{+0.3}_{-0.2}) \times 10^{-10}.
\end{equation}

Leptogenesis is a non-equilibrium process 
that takes place at temperatures $T\sim M_{1}$. 
This imposes an upper bound on the effective neutrino mass $%
\widetilde{m_{1}}$ given by the ``equilibrium neutrino mass'' 
\cite{Kolb:vq}, \cite{Fischler:1990gn}, \cite{Buchmuller:1992qc}: 
\begin{equation}
m_{*}=\frac{16\pi ^{5/2}}{3\sqrt{5}}g_{*}^{1/2}\frac{v^{2}}{M_{Pl}}\simeq
10^{-3}\ \mbox{eV}\; ,  \label{enm}
\end{equation}
where $M_{Pl}$ is the Planck mass ($M_{Pl}=1.2\times 10^{19}$ GeV). The sum
of all neutrino masses squared ${\overline{m}}^{2}$ is constrained, in the
case of normal hierarchy, to be below 0.20 eV \cite{Buchmuller:2002jk}. This
bound is automatically verified in the case $m_{1}=0$.

In the WB, where $M$ is diagonal, the product $m^{\dag }m$ is transformed, in
our framework, into: 
\begin{equation}
m^{\dag }m\rightarrow (F\cdot W_{R}\cdot K_{R})^{T}\ m^{\dag }m\ (F\cdot
W_{R}\cdot K_{R})^{*}  \; , \label{wbm}
\end{equation}
corresponding to the transformation: 
\begin{equation}
\nu _{R}^{0}\rightarrow (F\cdot W_{R}\cdot K_{R})^{*} \ \nu _{R}^{0} \; ,
\end{equation}
where $F$ denotes the matrix given in Eq. (\ref{fw}), $K_{R}$ is of the form 
$K_{R}=\mathrm{diag}
\left( -e^{i\pi /4},e^{i\pi /4},1\right)$ and $W_{R}$ is similar
to $W$ in Eq. (\ref{www}), 
with the masses of charged leptons replaced by the
masses of heavy neutrinos. Expanding $W_{R}$ up to the next order in the
ratios of masses leads to: 
\begin{equation}
W_{R} \simeq \left( 
\begin{array}{lll}
1+\frac{9i}{8}\frac{M_{1}}{M_{3}} & - \frac{M_{1}}{\sqrt{3}M_{2}}\left( 1+ 
\frac{i}{8}\frac{M_{2}}{M_{3}}\right) & -i\sqrt{\frac{2}{3}}\frac{M_{1}} {%
M_{3}}\left( 1+\frac{i}{4}\frac{M_{2}}{M_{3}}\right) \\ 
\frac{M_{1}}{\sqrt{3}M_{2}}\left( 1+\frac{3i}{4}\frac{M_{2}}{M_{3}}\right) & 
1+\frac{7i}{8}\frac{M_{2}}{M_{3}} & - \frac{i}{\sqrt{2}}\frac{M_{2}}{M_{3}}
\left( 1+i\frac{M_{2}}{M_{3}}\right) \\ 
-i\sqrt{\frac{3}{2}}\frac{M_{1}}{M_{3}} & -\frac{i}{\sqrt{2}}\frac{M_{2}} {%
M_{3}}\left( 1+\frac{i}{8}\frac{M_{2}}{M_{3}}\right) & 1+\frac{i}{4}\frac{%
M_{2}}{M_{3}}
\end{array}
\right) \; .
\end{equation}

From Eq. (\ref{wbm}) it is possible to show that the leading contributions
to the different matrix elements of ${m}^{\dag }m$, which appear in the
formulae relevant to leptogenesis are given by: 
\begin{eqnarray}
({m}^{\dag }m)_{11} &=&-\frac{2}{9}\ c_{D}^{2}\ \left( \frac{M_{1}}{M_{2}}%
\right) ^{2}\left[ a^{2}+b^{2}-9(a+b)\left( \frac{M_{2}}{M_{3}}\right)
\right]  \label{ldg1} \\
\mathrm{\mathrm{Re}}[({m}^{\dag }m)_{12}] &=&\frac{2}{3\sqrt{3}}\ c_{D}^{2}\
\left( \frac{M_{1}}{M_{2}}\right) \left[ a^{2}+b^{2}-6(a+b)\left( \frac{M_{2}%
}{M_{3}}\right) \right]  \label{ldg2} \\
\mathrm{Im}[({m}^{\dag }m)_{12}] &=&\frac{-5}{12\sqrt{3}}\ c_{D}^{2}\ \left( 
\frac{M_{1}}{M_{3}}\right) \left[ (a^{2}+b^{2})-\frac{24}{5}(a+b)\left( 
\frac{M_{2}}{M_{3}}\right) \right]  \label{ldg3} \\
({m}^{\dag }m)_{13} =e^{-i\pi /4}&\sqrt{\frac{2}{3}}&\ c_{D}^{2}\ \left( 
\frac{M_{1}}{M_{2}}\right) \left[ (a+b)-\frac{27}{2}\left( \frac{M_{2}}{M_{3}%
}\right) -\frac{i}{6}(a^{2}+b^{2})\right].  \label{ldg4}
\end{eqnarray}

As an example, for $M_{1}\sim 10^{10}$ GeV and $M_{2}\sim 10^{13}$ 
GeV, which implies fixing the parameters 
$A$ and $B$, together with $M_{3}$, which is fixed
to be of order $10^{16}\ \mathrm{GeV}$ as explained in Section 2, we obtain: 
\begin{equation}
\widetilde{m_{1}}\sim \ 10^{-5}\ \mbox{eV},\qquad 
\epsilon _{N_{1}}\sim 10^{-7}\; ,
\end{equation}
where we have chosen $a=0.052998$, $b=0.102998$ and $B=0.01$, as in the
explicit example of the previous section, together with $A=6\times 10^{-6}$
and $c_{R}=3\times 10^{24}$ eV. This leads to the observed baryon asymmetry
of the Universe, assuming a washout factor of order $10^{-1}$.

Expressions (\ref{ldg1}) to (\ref{ldg4}) can be
rewritten in terms of initial parameters $A, B$ 
and $a, b$ with the help of Eq. (\ref{abg}). 
Since $A$ is not constrained by the low energy physics, there is
still a free variable parametrizing leptogenesis. Varying $A$ corresponds to
different choices of $M_{1}$, which, as was pointed out in the case of
hierarchy of heavy neutrino masses, is the most important of the three
masses. On the other hand, low energy physics fixes the ratios 
$\frac{a}{\sqrt{B}}$, $\frac{b}{\sqrt{B}}$ for variable $B$.

\section{Conclusions}

We have investigated the pattern of lepton masses and mixing within the
framework of USY, where all Yukawa couplings have equal modulus, the
flavour structure being all contained in the phases of those couplings.

In our scheme, three of the four parameters of the charged lepton mass
matrix $m_{l}$ are related to the three charged lepton masses, the
fourth one (the phase $\theta $) giving the dominant contribution to $\sin
^{2}(2\theta _{\mathrm{atm}})$; the three parameters in the Dirac neutrino
mass matrix $m$ together with $c_{R}$ are constrained by the two neutrino
squared mass differences (for each choice of $B$) and by $\sin ^{2}(2\theta
_{\mathrm{sol}})$. Furthermore, the assumption that $c_{D}$ is of the order
of the weak scale fixes the scale of the heavy neutrino masses and, with our
choice of hierarchical heavy neutrino masses, it determines, to an excellent
approximation, the mass of the heaviest neutrino. 
The remaining parameters $A$
and $B$ are dependent on the hierarchy of the heavy neutrino masses, for
which there is no experimental information available. It should be noticed
that $A$ has no influence on low energy physics. Strong hierarchy of heavy
neutrino masses requires $A$ and $B$ to be much smaller than $1$.
Hierarchical heavy neutrinos are generally favoured in scenarios of
baryogenesis through leptogenesis (together with one light neutrino
with zero mass), since in this case 
the washout processes tend to be less important.

In conclusion, we have suggested an ansatz for the leptonic flavour
structure, where all the experimental data on lepton masses and mixing are
reproduced and specific correlations between measurable
quantities are obtained,
such as $U_{13}$, the ratio of neutrino squared mass differences,
the strength of leptonic Dirac-type CP violation and the mass parameter 
$\mid \langle m \rangle \mid $, 
measurable in neutrinoless double beta decay. In particular,
in the scheme we propose, it is possible to have $U_{13}$ close to the
present experimental bound together with a large value for $I$, provided the
ratio $r\equiv (\Delta m_{21}^{2}$/$\Delta m_{32}^{2})^{1/2}$ lies within
the lower values of the experimentally allowed range. In this case the
observation of leptonic CP violation due to a Dirac-type CP violating phase
would be within the reach of future neutrino experiments. Finally, it was
shown that in this scheme a sufficient amount of  
baryon asymmetry of the Universe can be generated through leptogenesis.

\section*{Acknowledgements}

GCB and MNR thank the Theory Division of CERN for warm hospitality. This
work was partially supported by Funda\c{c}\~{a}o para a Ci\^{e}ncia e a
Tecnologia (FCT, Portugal), through the projects CERN/FIS/43793/2001,
POCTI/36288/FIS/ 2000, CFIF - Plurianual (2/91), which  
is partially funded through POCTI (FEDER),
and by CERN.

\newpage 
\begin{figure}[t]
\vspace{-2.0truecm} \centerline{\ \epsfysize=8.0truecm \epsffile{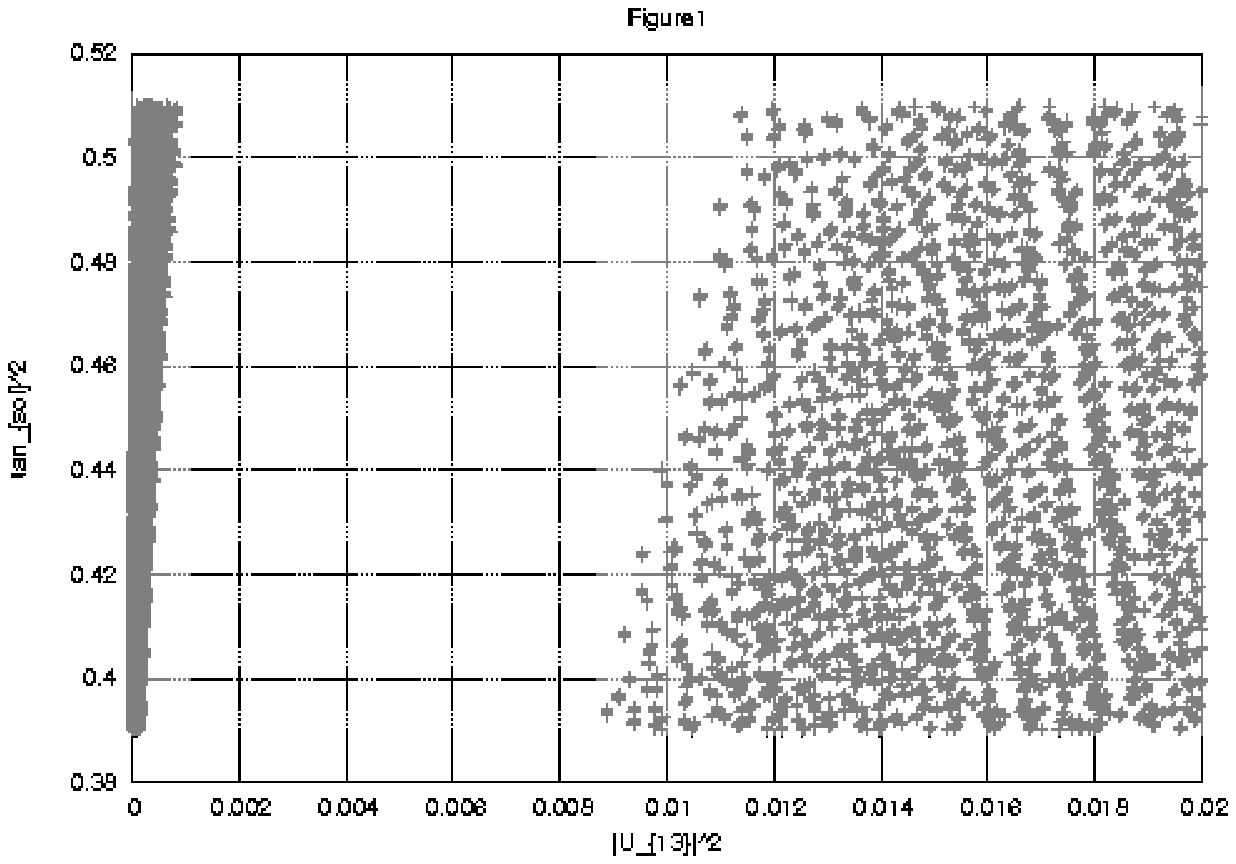}}
\caption{Region allowed for $|U_{13}|^{2}$ and $\tan^{2}(\theta_{sol})$.}
\label{figsv}
\end{figure}

\begin{figure}[t]
\vspace{2.0truecm} \centerline{\ \epsfysize=8.0truecm \epsffile{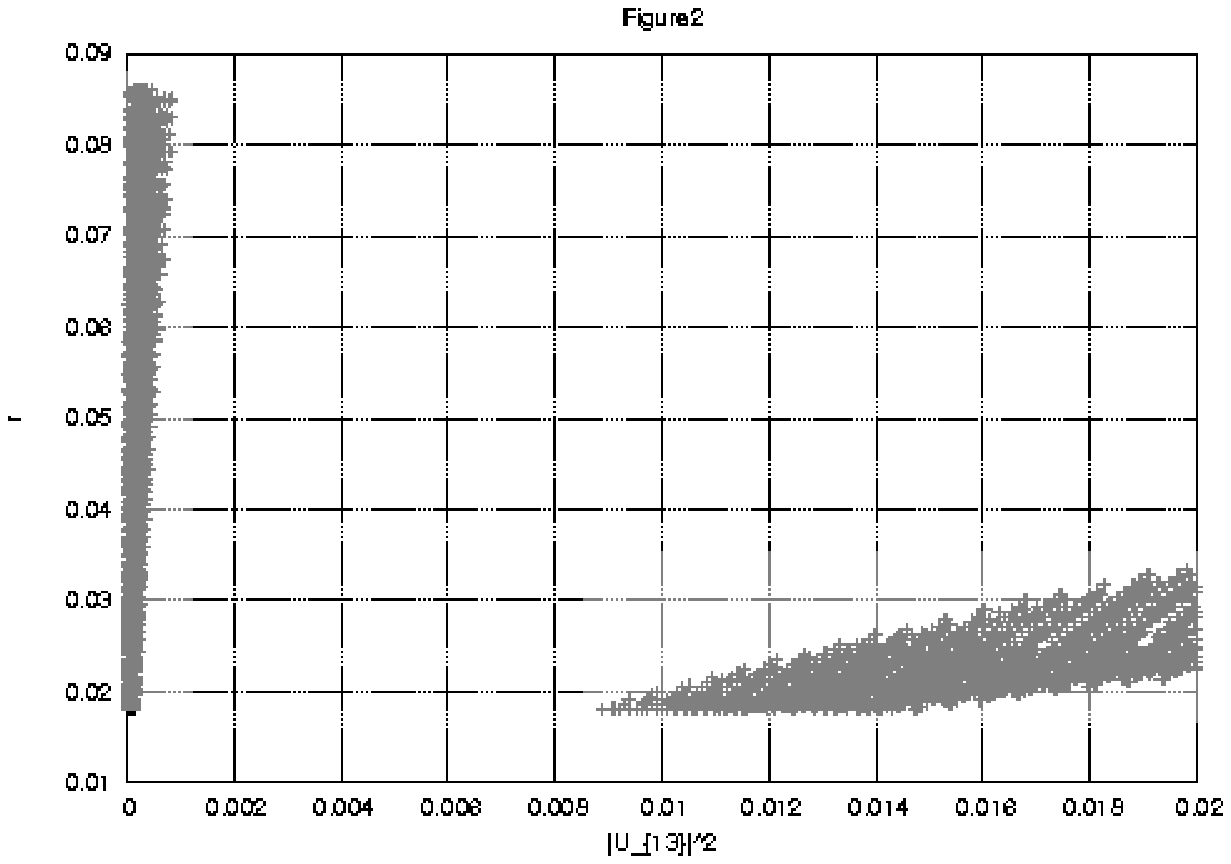}}
\caption{Region allowed for $|U_{13}|^{2}$ and the mass ratio $r=(\Delta
m^{2}_{21}/\Delta m^{2}_{23})^{(1/2)}$.}
\label{figvm}
\end{figure}

\newpage

\begin{figure}[t]
\vspace{-2.0truecm} \centerline{\ \epsfysize=8.0truecm \epsffile{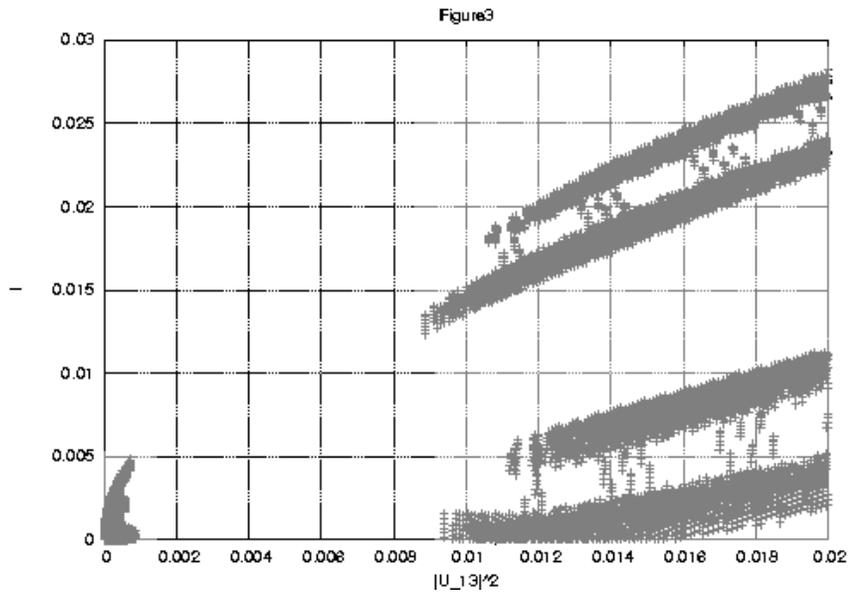}}
\caption{Region allowed for $|U_{13}|^{2}$ and the invariant $I$.}
\label{figvj}
\end{figure}

\begin{figure}[t]
\vspace{2.0truecm} \centerline{\ \epsfysize=8.0truecm \epsffile{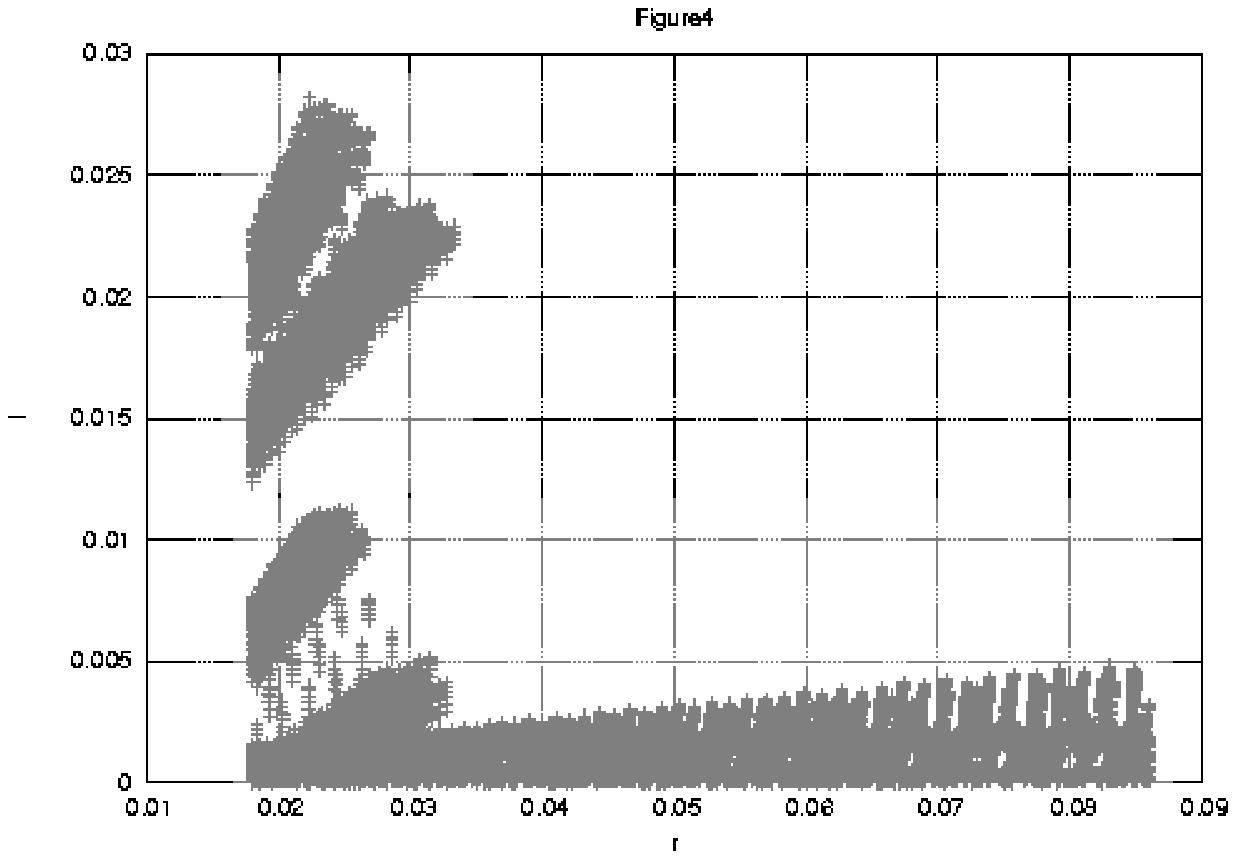}}
\caption{Region allowed for the mass ratio $r=(\Delta m^{2}_{21}/\Delta
m^{2}_{23})^{(1/2)}$ and $I$.}
\label{figmj}
\end{figure}
\newpage

\begin{figure}[t]
\vspace{-2.0truecm} \centerline{\ \epsfysize=8.0truecm \epsffile{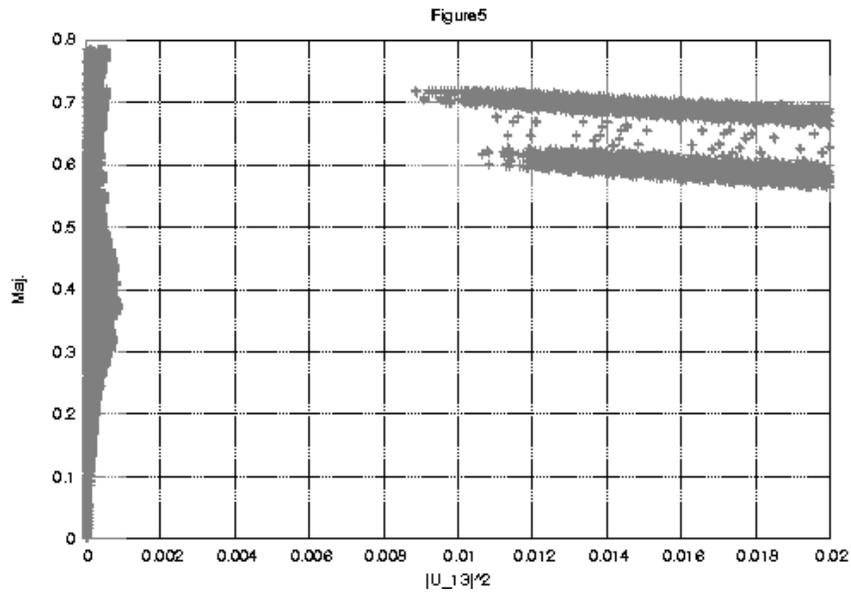}}
\caption{Region allowed for $|U_{13}|^{2}$ and the Majorana phase
combination $\delta-\alpha$.}
\label{figmajo}
\end{figure}

\end{document}